\documentclass[12pt,a4paper]{article}
\RequirePackage{lineno}
\usepackage{epsfig}
\usepackage{graphicx}                  
\usepackage{siunitx}
\usepackage{ae}
\usepackage{amsmath}
\usepackage{amssymb}
\usepackage{graphics}
\usepackage{dcolumn}
\usepackage{bm}
\usepackage[symbol]{footmisc}
\usepackage{setspace}
\usepackage{ragged2e}
\usepackage{wrapfig}
\usepackage[left=.4in, right=.4in, top=0.8in]{geometry}
\usepackage[margin=0.4in,font=small,labelfont=bf]{caption}
\begin{document}
\begin{center}

{\Large
			{\bf Cosmic ray flux and lockdown due to COVID19 in Kolkata - any correlation?}}
		\vskip 0.8 true cm
		\renewcommand{\thefootnote}{\fnsymbol{footnote}}
		
		{\bf A. Sen,
			S.~Chatterjee, S.~Roy,  R.~Biswas, S.~Das, S.~K.~Ghosh, and S.~Biswas$\footnote[1]{Corresponding author}$$ \let\thefootnote\relax\footnote{e-mail: saikat@jcbose.ac.in}$}
		
		{\it  Department of Physics and Centre for Astroparticle Physics and Space Science, Bose Institute, \\ EN-80, Sector V, Kolkata-700091, INDIA}
		
		\end{center}

	\justify

\abstract
Cosmic ray muon flux is measured by the coincidence technique using plastic scintillation detectors in the High Energy Physics detector laboratory at Bose Institute, Kolkata. Due to the COVID19 outbreak and nationwide complete lockdown, the laboratory was closed from the end of March 2020 till the end of May 2020. After lockdown, although the city is not in its normal state, we still were able to take data on some days. The lockdown imposed a strict restriction on the transport service other than the emergency ones and also most of the industries were shut down in and around the city. This lockdown has significant effect on the atmospheric conditions in terms of change in the concentration of air pollutants. We have measured the cosmic ray flux before and after the lockdown to observe the apparent change if any due to change in the atmospheric conditions. In this article, we report the measured cosmic ray flux at Kolkata (22.58$^{\circ}$N 88.42$^{\circ}$E and  11~m Above Sea Level) along with the major air pollutants present in the atmosphere before and after the lockdown.

		
\section{Introduction}
\label{intro}
		
Cosmic ray consists of high energy particles that mostly originate from outer space, with some very high energy particles even thought to have an extragalactic origin. Primary cosmic rays consist of 90\% protons, 9\% alpha particles and of other heavier nuclei~\cite{cosmic_ray}. This primary cosmic rays interact with the gas molecules in the atmosphere and produce secondary cosmic rays. These secondary particles consist mostly of pions and some kaons. Neutral pions~($\pi^0$) decay into gamma rays that generate electromagnetic showers (e$^+$, e$^-$, $\gamma$), which posses low penetration power. Charged pions~($\pi^+$, $\pi^-$) decay into muons and neutrinos. Neutrinos have a very small cross-section for interaction and typically pass through the earth without any further interactions. On the other hand, muons are heavy particles and thus loss of energy through bremsstrahlung is negligible for them. This makes the muon a very penetrating particle, unlike electron. The muon has a lifetime of 2.2~$\mu$s  yet it still makes it down to detectors at the surface of the earth traversing through the atmosphere. This is because the muon travels at a speed that is close to that of light and thus experiences relativistic time dilation and therefore can be detected by our detectors. Since the secondary cosmic rays are mostly muons and they can travel large distance through the atmosphere before they are detected, it will be really interesting if any correlation of this cosmic ray muon flux with the change in atmosphere in terms of the concentration of air pollutants is found~\cite{S. Jain, S. Chen, A. Chatterjee}. 
		
For this study, cosmic ray flux has been measured in our laboratory using plastic scintillator detectors before and after the imposition of lockdown due to the COVID19 pandemic. The effects of atmospheric pressure and temperature on the muon flux has also been studied here.  A brief description of the change of atmospheric parameters due to lockdown is discussed in the next section. The succeeding sections consist the details of the experimental setup, results followed by summary and discussions.

\section{Effect of lockdown}
\label{lockdown}

\begin{figure}[htb!]
\centering
\includegraphics[scale=0.65]{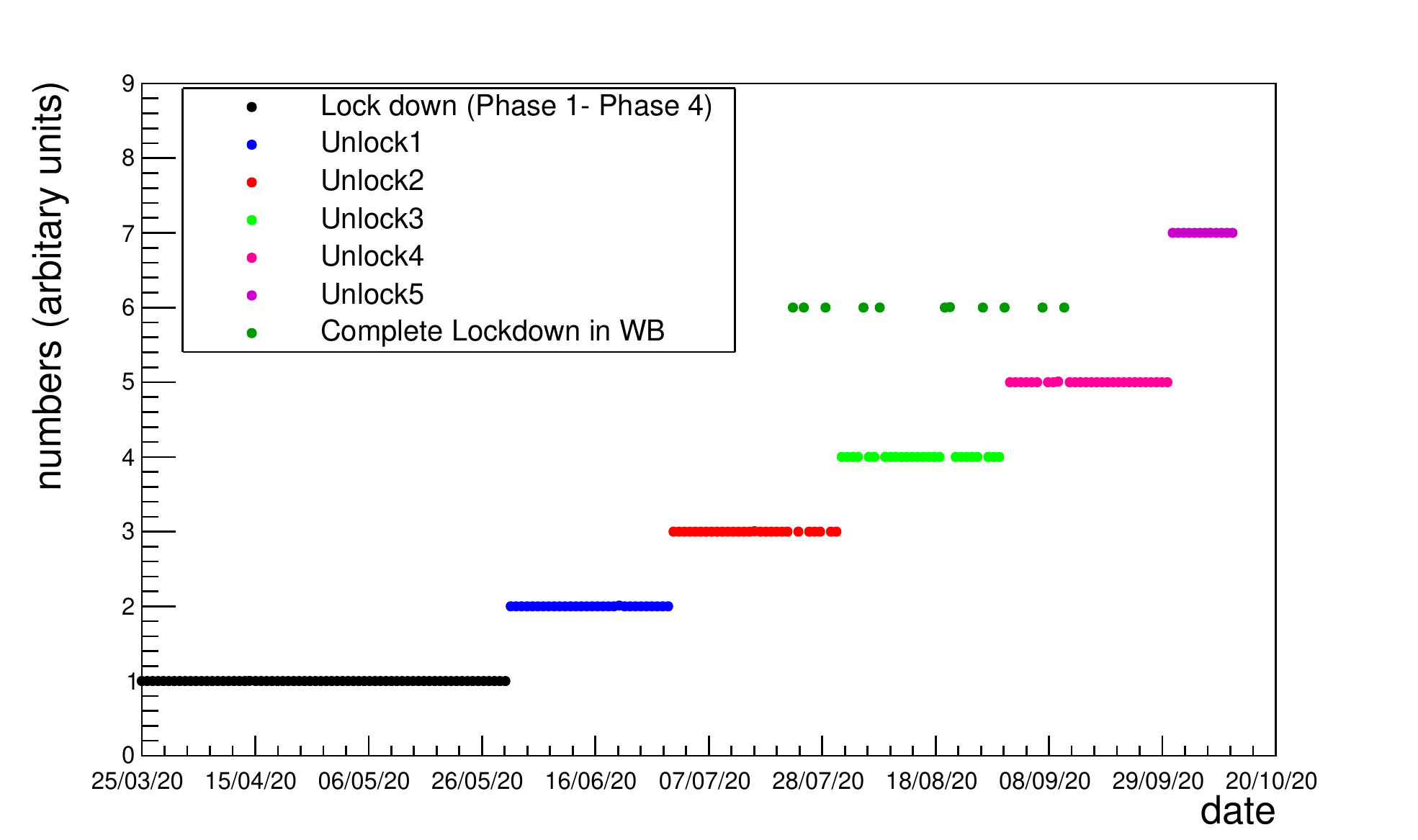}
\captionof{figure}{(Colour online) Different phases of lockdown and unlock as a function of date.}
\label{fig1}
\end{figure}

India is at a critical stage in its fight against COVID19 with positive cases crossing 71,73,565  and death toll at 1,09,894 until October 12, 2020~\cite{covid}. The entire country was under complete lockdown from March~25 to April~14, 2020, for 21 days, which was further extended by the Government of India until May 3, 2020, followed by the third phase of lockdown till May~17, 2020, and the fourth phase till May~31, 2020, to tackle the spread of COVID19. Restrictions on social gathering and travelling resulted in the shutdown of all the businesses which include industries, transport (air, water, and surface), markets, shops, tourism, construction and demolition, hotels and restaurants, mining and quarrying, etc. except the essential services like groceries, milk, medicines and emergency services like hospital, fire service and administration. In June 2020 Governments, both state and central, declared
restricted unlocking phase. While Unlock phases were started from 1$^{st}$ June 2020, there were complete lockdown in West Bengal on some selected dates to combat against COVID19. The dates of lockdown are mentioned in Fig.~\ref{fig1}.
		
The complete lockdown is marked with 1, different unlock phases are marked as 2, 3 so on and the complete lockdown days in West Bengal during the unlock phases are marked with 6. During the lockdown~(25$^{th}$ March to 6$^{th}$ April 2020; Lockdown Phase-1) and before lockdown~(10$^{th}$ - 20$^{th}$ March 2020), significant variation in the concentrations of the five most abundant pollutants in the air~(PM$_{2.5}$, PM$_{10}$, NO$_{2}$, CO, O$_3$) are observed in Kolkata. The concentrations of air pollutants in Kolkata are decreased by $\sim$23\% (PM$_{2.5}$), $\sim$34\% (PM$_{10}$), $\sim$60\% (NO$_{2}$), $\sim$29\% (CO) while the $O_3$ concentration is increased by $\sim$17\% due to the clearer atmosphere as compared to that before lockdown period~\cite{S. Jain}. We also looked into the last year data for the same period (March-April, 2019) and found that, during the lockdown, the concentration of air pollutants are decreased by $\sim$27\% (PM$_{2.5}$), $\sim$32\% (PM$_{10}$), $\sim$66\% (NO$_{2}$), $\sim$16\% (CO) and $O_3$ concentration is increased by $\sim$87\%~\cite{S. Jain}.
		
We have used the live day to day data from Ref~\cite{pollution} of the concentrations of the seven major air pollutants and studied the effects of them on cosmic ray flux. In our work, we have reported the measured muon fluxes before and after the lockdown at Kolkata and tried to correlate the same with the change in the concentrations not only of the individual components of air pollutants but also the total amount of pollutants.
		
\section{Experimental Setup}
\label{setup}
		
\begin{figure}[htb!]
\centering
\includegraphics[scale=0.5]{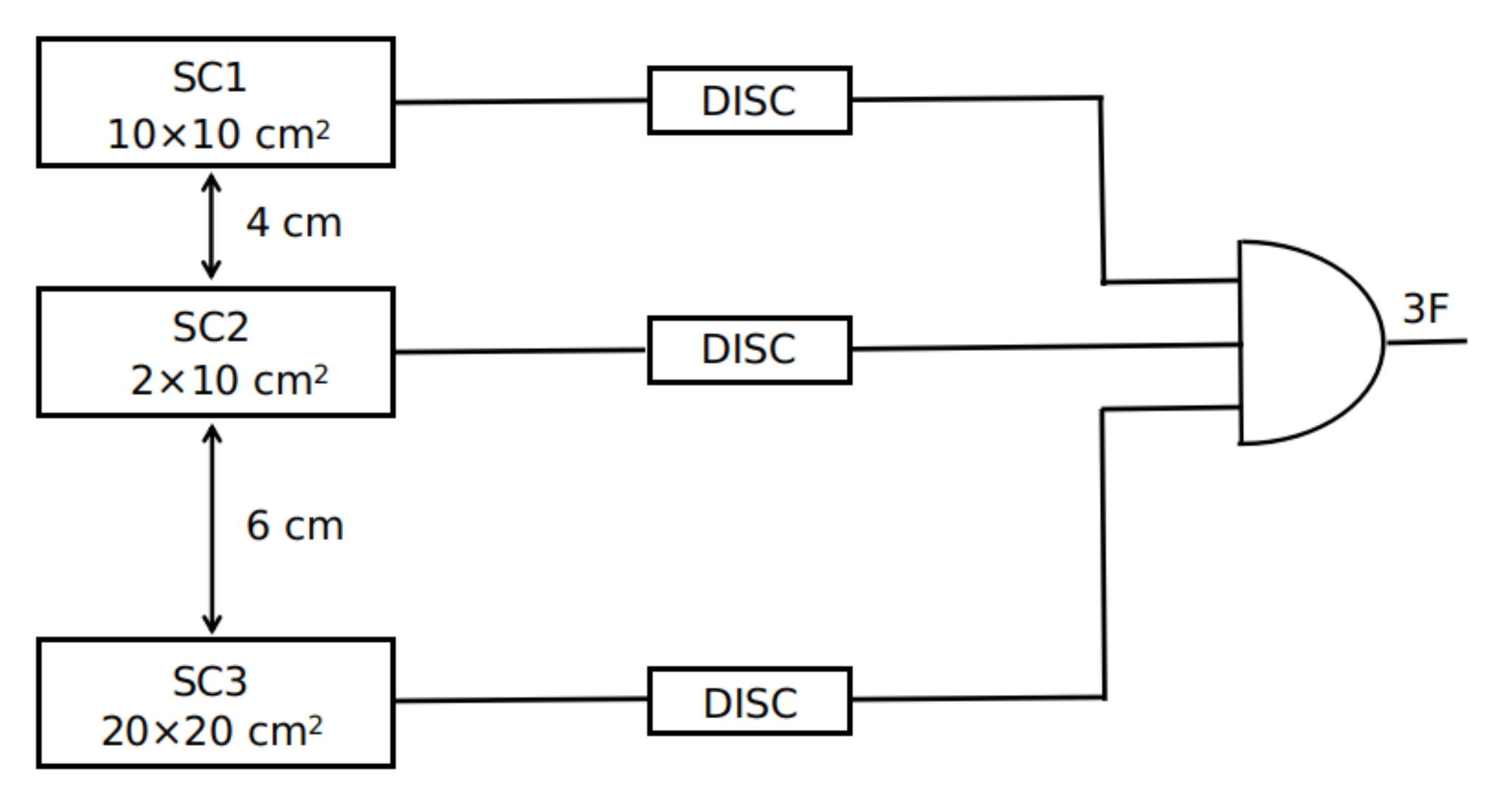}
\captionof{figure}{Schematic of the experimental setup for muon flux measurement at the laboratory}
\label{fig2}
\end{figure}

The schematic of the muon flux measurement setup is shown in Fig.~\ref{fig2}. Three plastic scintillators tagged as SC1, SC2, and SC3, made by using BC400 material are used in this setup~\cite{S. Roy, S. Shaw}. The dimensions of these scintillators are 10$\times$10~cm$^2$ , 2$\times$10~cm$^2$ and  20$\times$20~cm$^2$ respectively. The coincidence area of three detectors is 20~cm$^2$. The distance between the top and bottom scintillator is $\sim$10~cm whereas that between the top and the middle one is 4~cm. Each scintillator is connected with a Photo Multiplier Tube~(PMT) and a base where one SHV and one BNC connectors are provided for the application of High Voltage~(HV) and collection of signals respectively. +1550~V are applied to all the PMTs. Thresholds to the discriminators are set to -15 mV for all the scintillators. The width of each discriminator output is kept at 50~ns. The coincidence of these three signals is achieved using a logic unit. The three-fold coincidence signal is then counted using a scaler and then divided by the product of the area of coincidence window~(20~cm$^2$), muon detection efficiency of the system~($\sim$72\%)~\cite{S. Shaw} and the measurement time to get the muon flux. Each data point represents a 30~minutes long measurement. To check the health of the individual detectors, the singles count rate of all the modules is measured several times. It is found that the singles count rate of scintillators SC1, SC2, and SC3 are found to be $\sim$77, $\sim$28, and $\sim$171 Hz respectively.
		
\section{Results}
\label{res}

\begin{figure}[htb!]
\centering
\includegraphics[scale=0.65]{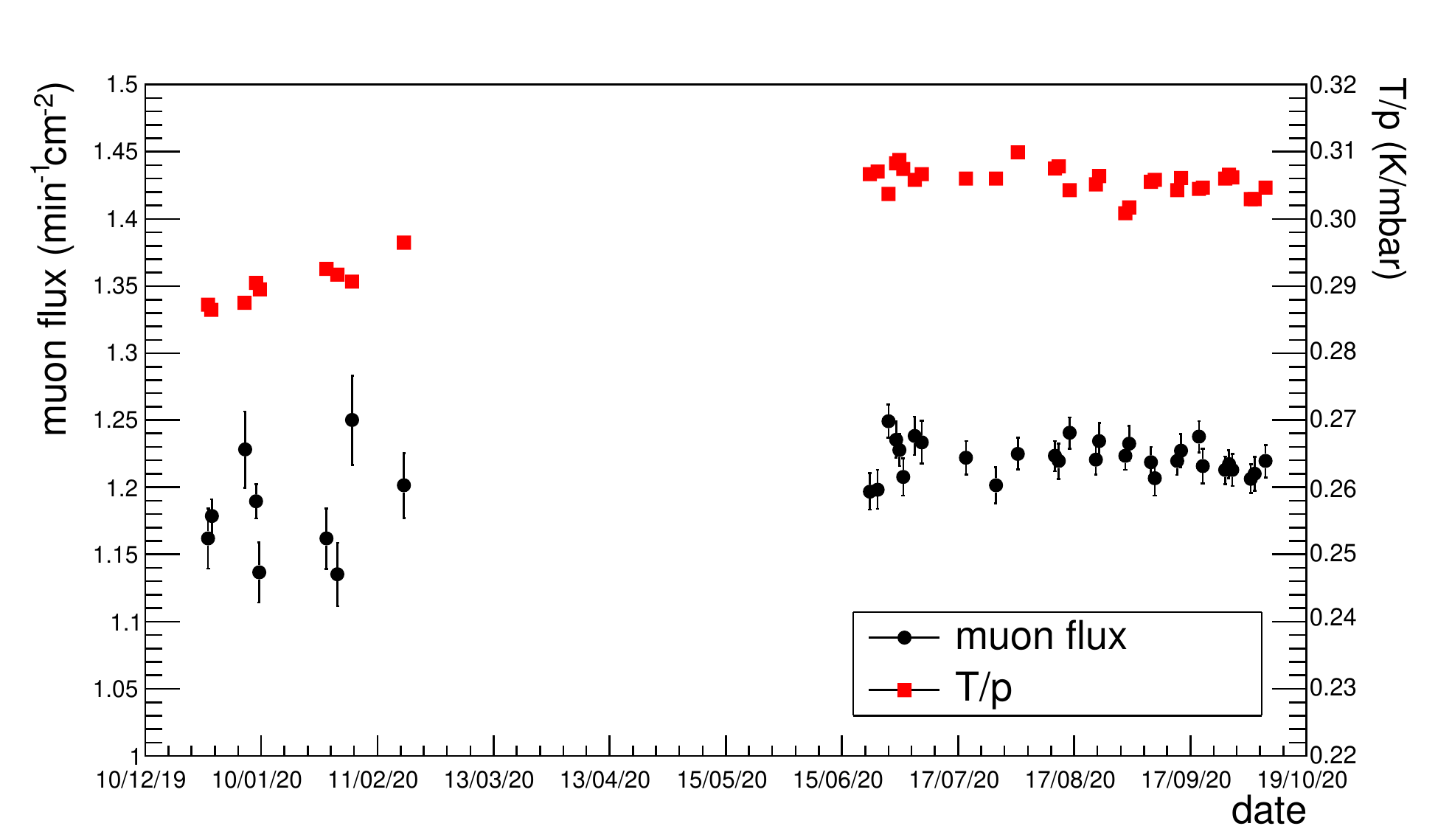}
\captionof{figure}{(Colour online) Cosmic ray muon flux and T/p as a function of date.}
\label{flux_date}
\end{figure}
\begin{figure}[htb!]
\centering
\includegraphics[scale=0.65]{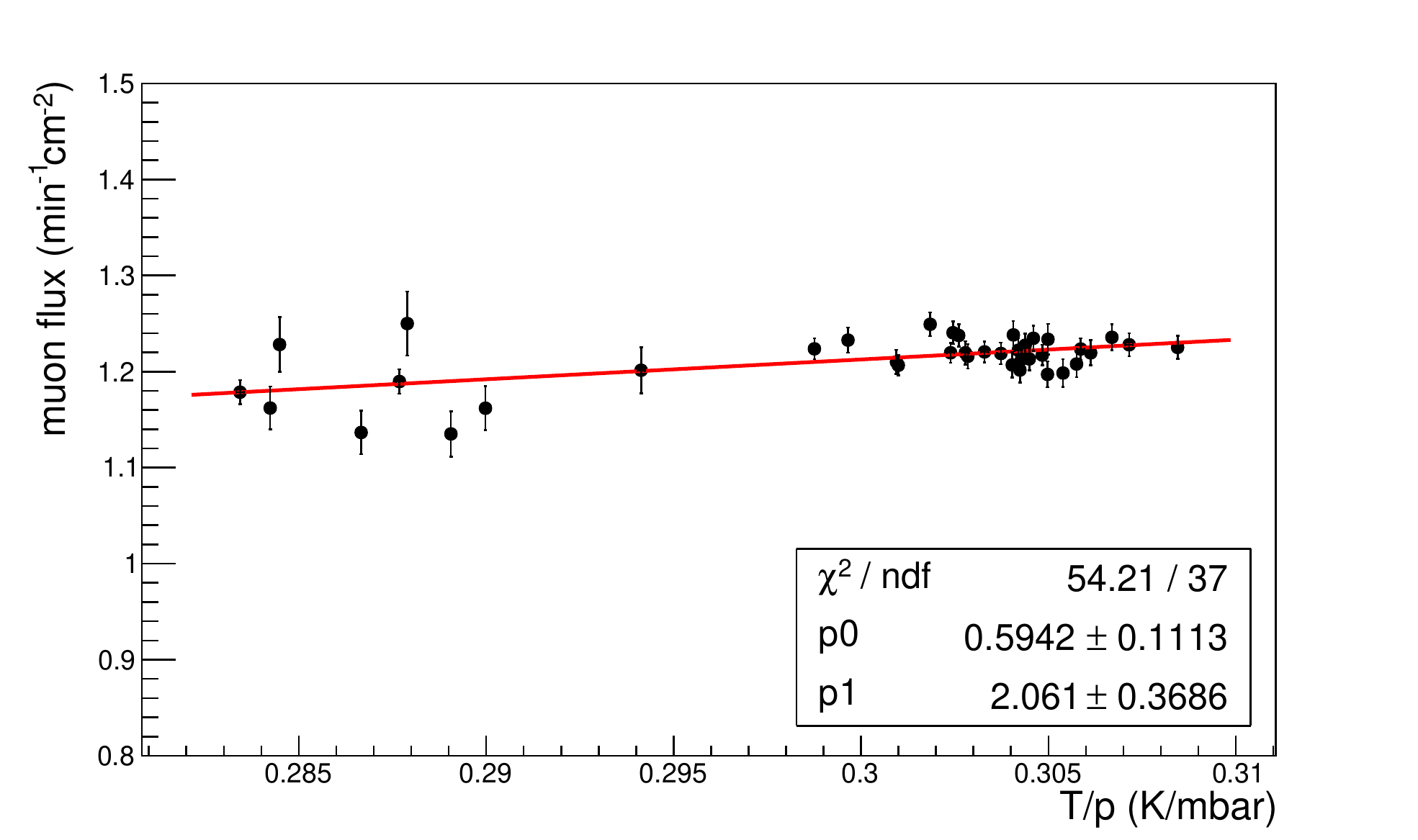}
\captionof{figure}{(Colour online) Correlation of cosmic muon flux with the ratio of temperature by pressure}
\label{fig3}
\end{figure}
\begin{figure}[htb!]
\centering
\includegraphics[scale=0.65]{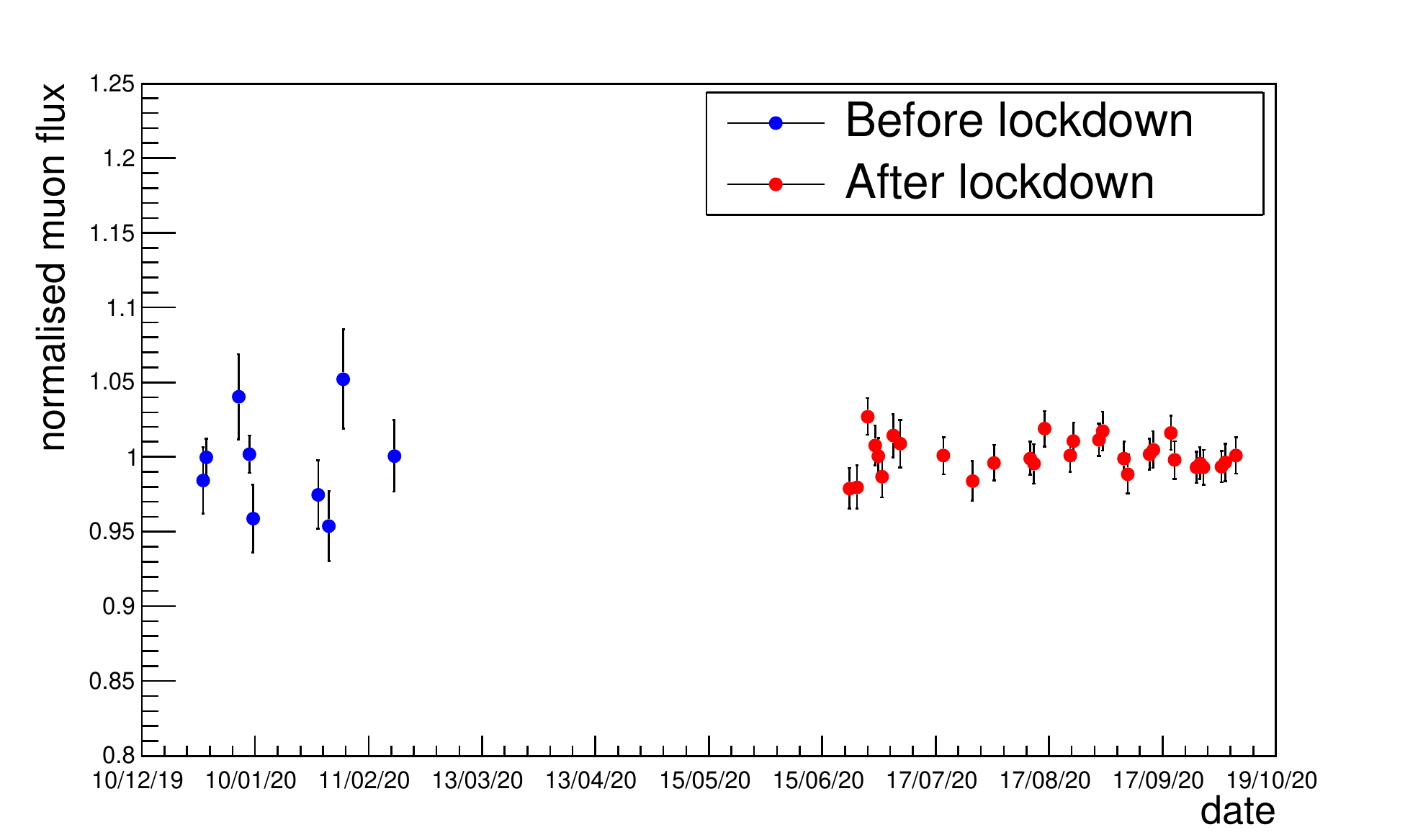}
\captionof{figure}{(Colour online) Normalised muon flux as a function of date. The gap between the two sets of data is the period of lockdown.}
\label{fig4}
\end{figure}
\begin{figure}[htb!]
\centering
\includegraphics[scale=0.45]{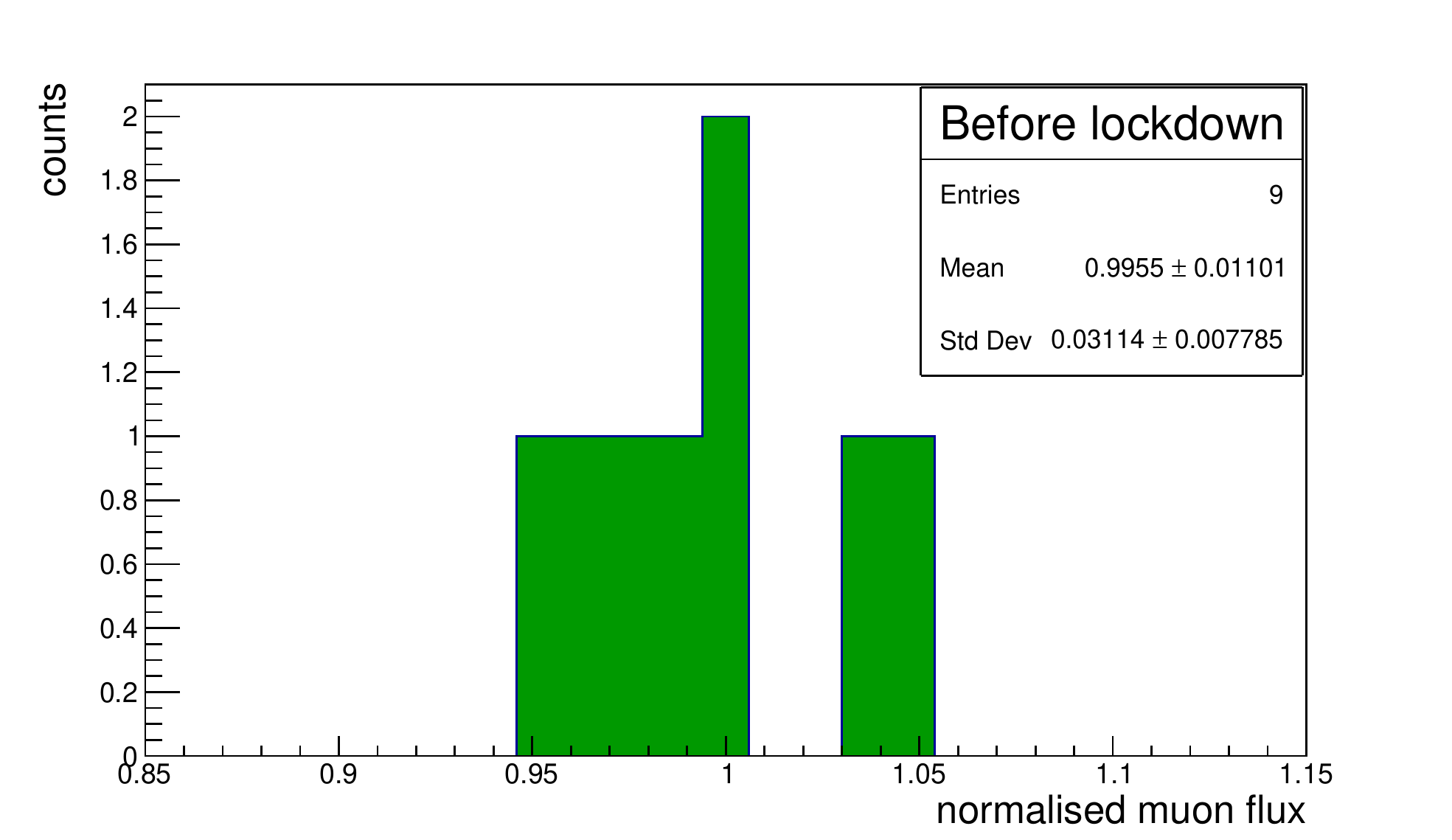}
\includegraphics[scale=0.45]{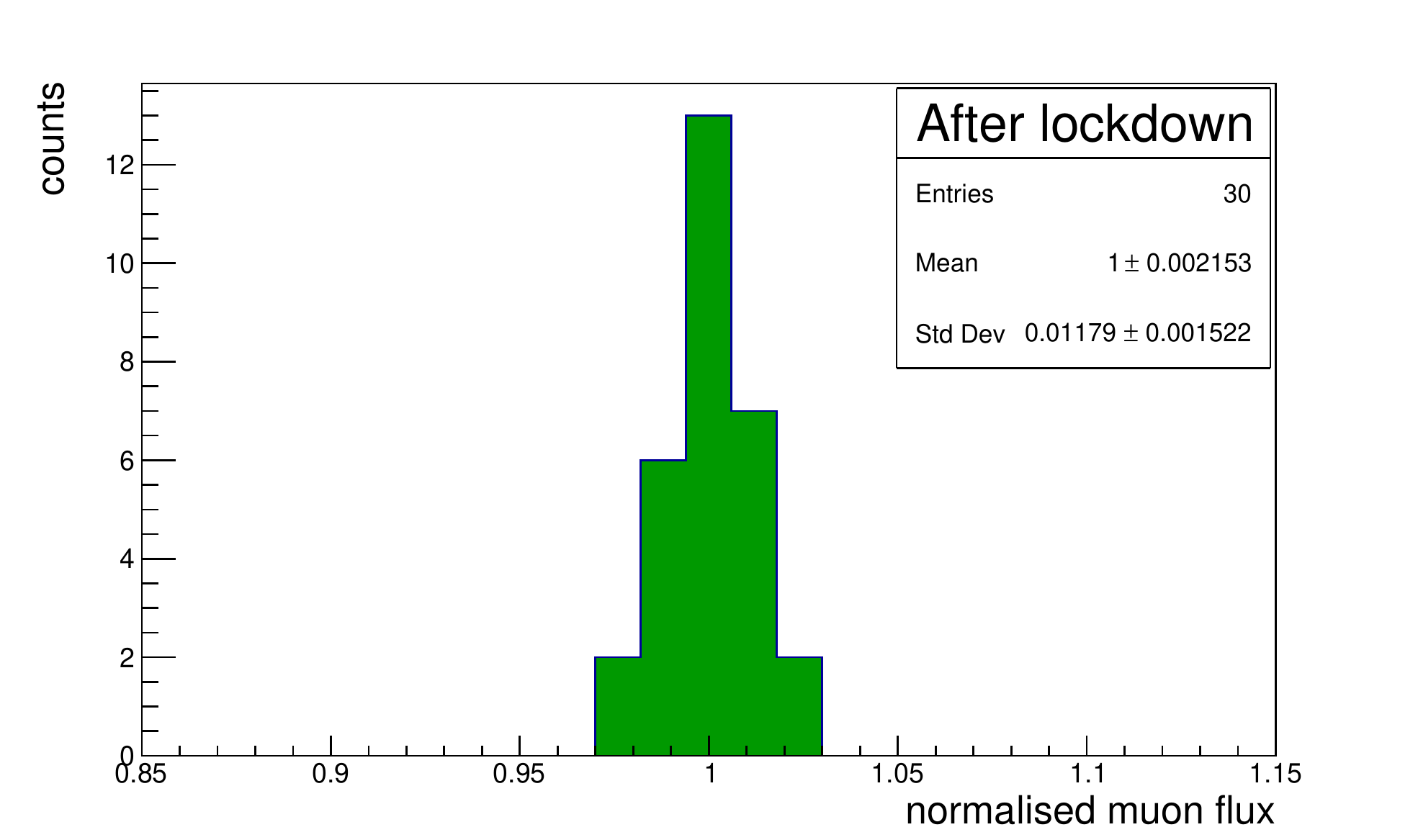}
\captionof{figure}{(Colour online) Distribution of normalised muon flux before and after lockdown}
\label{fig5}
\end{figure}

The cosmic ray flux is measured in Kolkata before and after the lockdown due to the COVID19 outbreak. The average muon flux before and after the lockdown is shown in Fig~\ref{flux_date} as a function of time. We had a very small amount of data before lockdown. The cosmic ray flux has a dependence on atmospheric parameters like temperature and pressure~\cite{M. Neira, M. Zazyan}. In this work, the temperature and pressure data are collected from Ref~\cite{tp}. The ratio of temperature and pressure as a function of day is also shown in Fig~\ref{flux_date}. In order to normalise the temperature~$\it(T=t+273)$ and pressure~$\it(p)$ effects, a simple correlation between cosmic muon flux and $\it T/p$ is studied using a relation of the form $\it p0+p1(T/p)$. The correlation of muon flux and $\it T/p$ is shown in Fig.~\ref{fig3}. The parameters obtained from the correlation are {\it 0.59~$\pm$~0.11~min$^{-1}$cm$^{-2}$~(p0)} and {\it 2.06~$\pm$~0.37~min$^{-1}$cm$^{-2}$k$^{-1}$mbar~(p1)} respectively. 

A positive correlation is observed between the muon flux and $\it T/p$. Using the parameters, the muon flux measured before and after the lockdown is normalised and shown in Fig~\ref{fig4}.

In Fig.~\ref{fig5}, the distribution of the normalised muon flux is shown before and after the lockdown. It is found that the mean normalised muon flux before the lockdown period is 0.996 with a standard deviation of 0.031, whereas that after lockdown is 1.00 with a standard deviation of 0.012 i.e. 0.4~\% increment in muon flux is found after lockdown. 

\begin{figure}[htb!]
\centering
\includegraphics[scale=0.85]{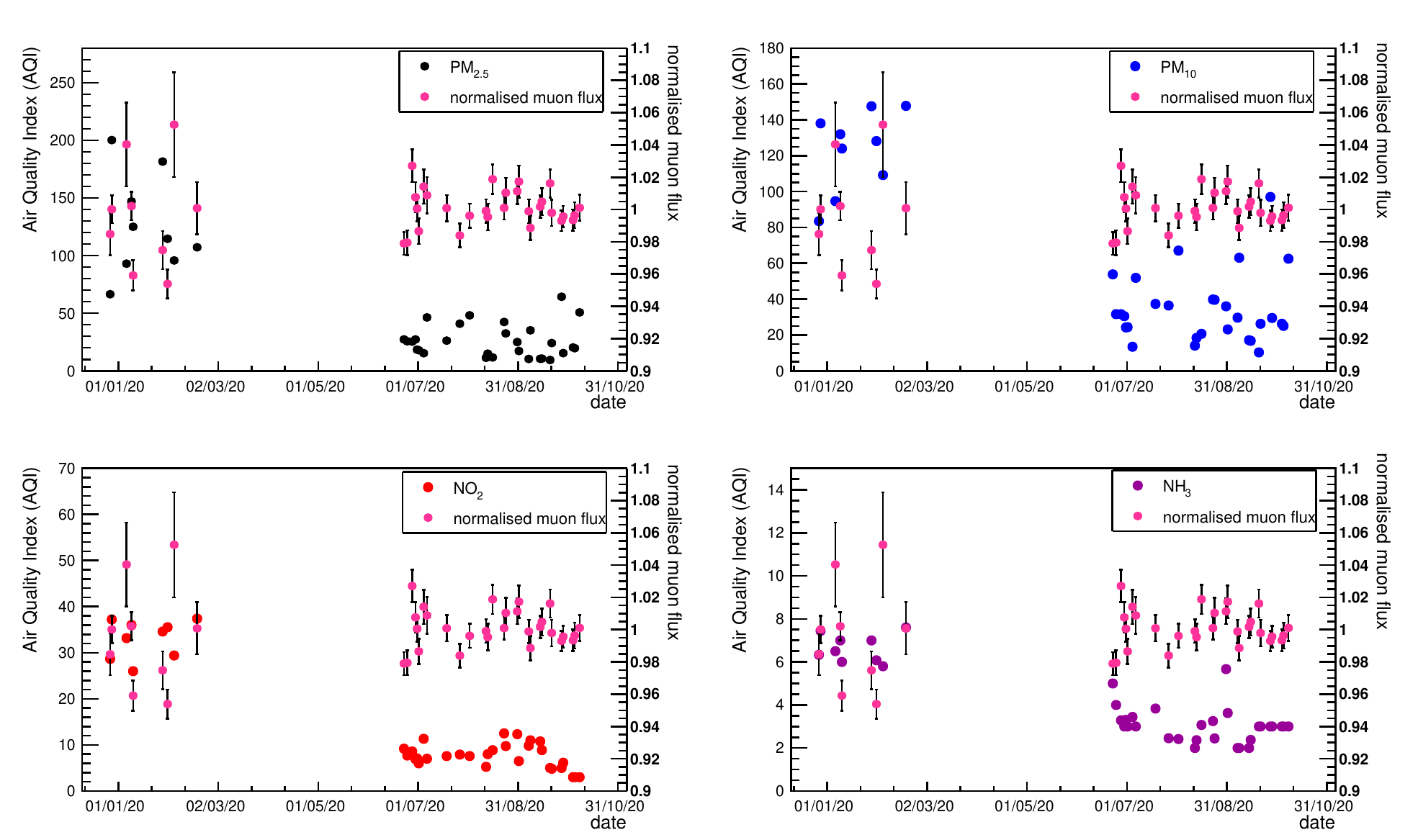}
\includegraphics[scale=0.85]{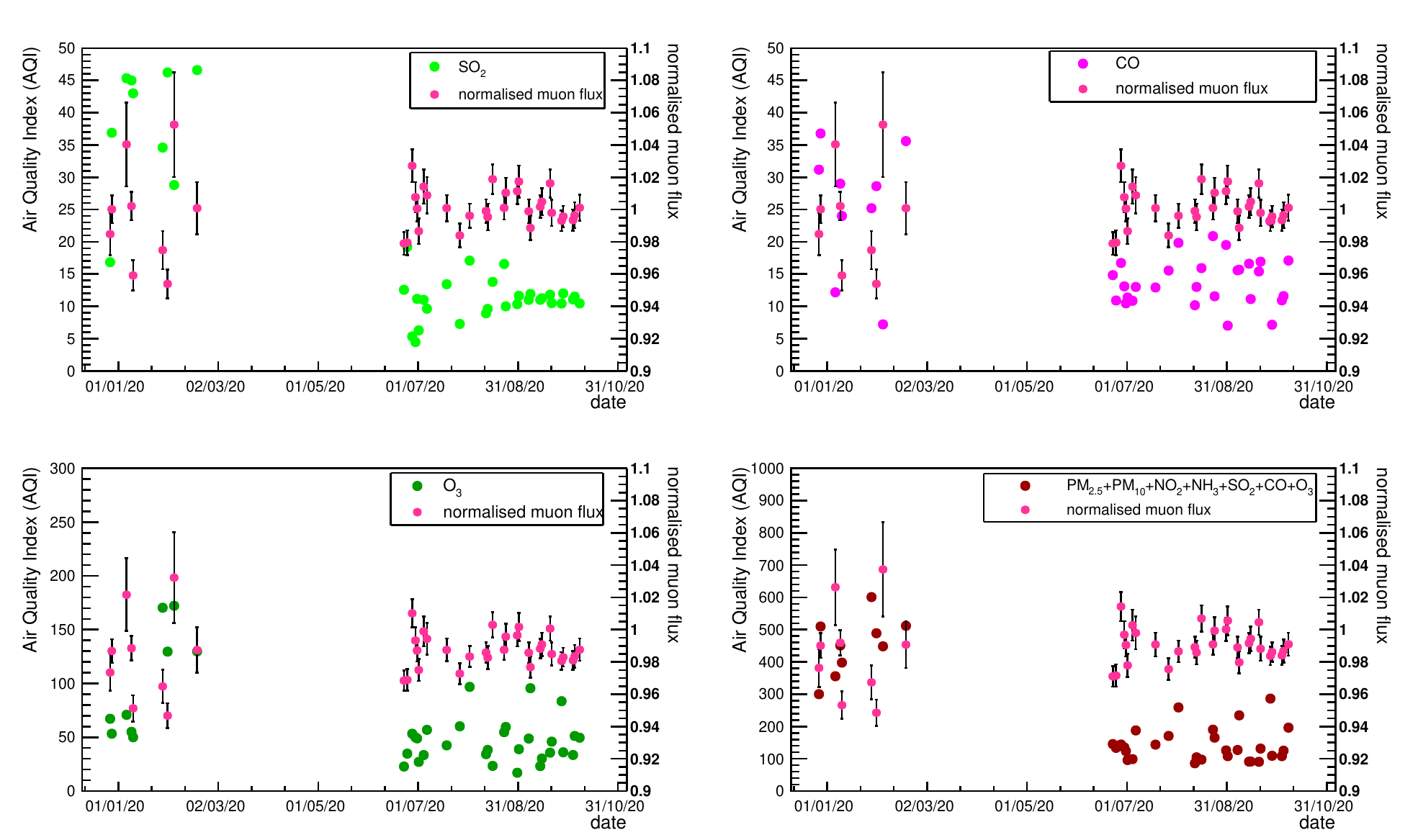}
\captionof{figure}{(Colour online) Average Air quality index (AQI) of 7 most abundant air pollutants measured at Bidhannagar, Kolkata station \cite{pollution} and the normalised muon flux as a function of date. }
\label{aqi_date}
\end{figure}

\begin{figure}[htb!]
\centering
\includegraphics[scale=0.85]{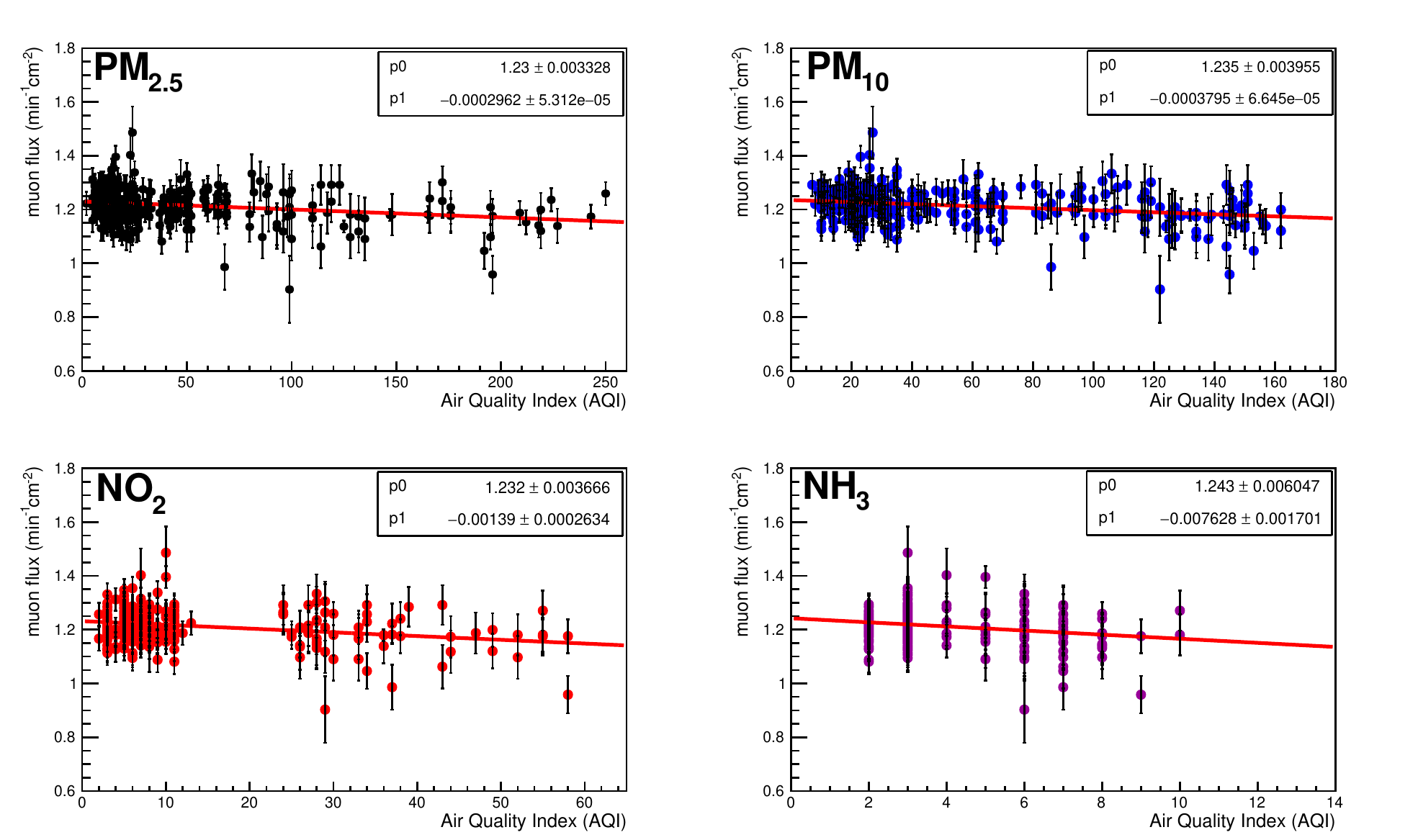}
\includegraphics[scale=0.85]{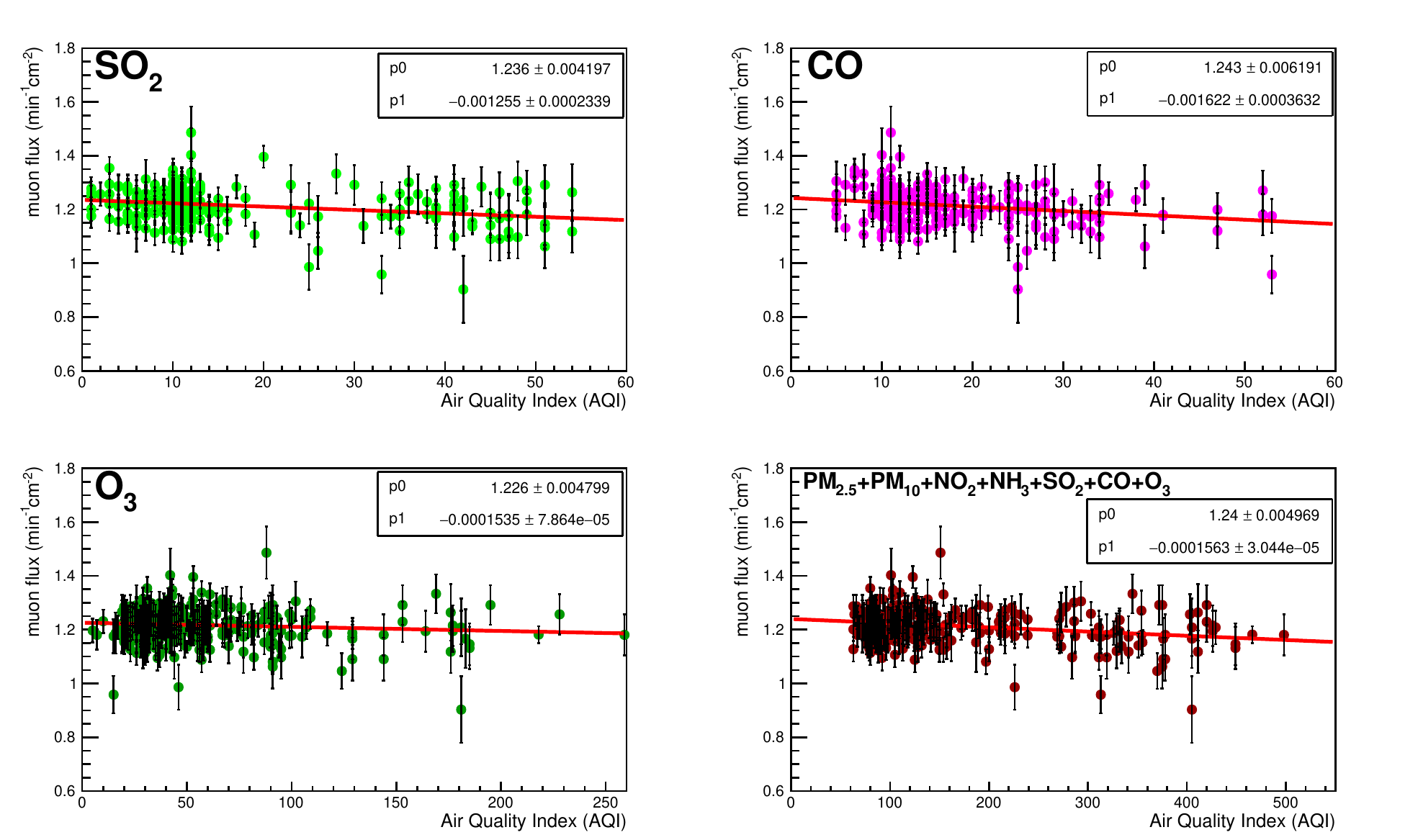}
\captionof{figure}{(Colour online) Measured muon flux as a function of the air quality index (AQI) of 7 most abundant air pollutants.}
\label{flux_aqi}
\end{figure}

\begin{figure}[htb!]
\centering
\includegraphics[scale=0.85]{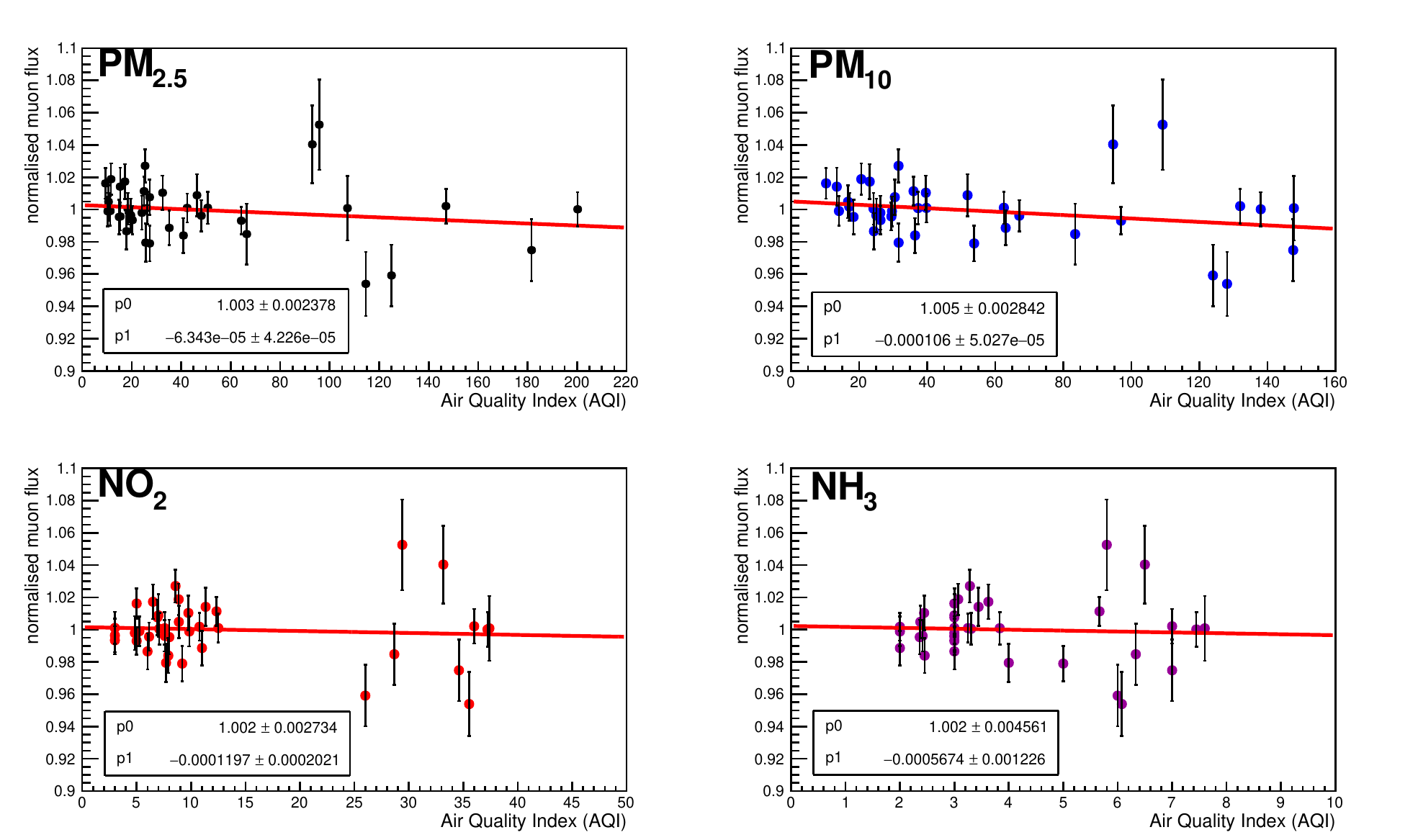}
\includegraphics[scale=0.85]{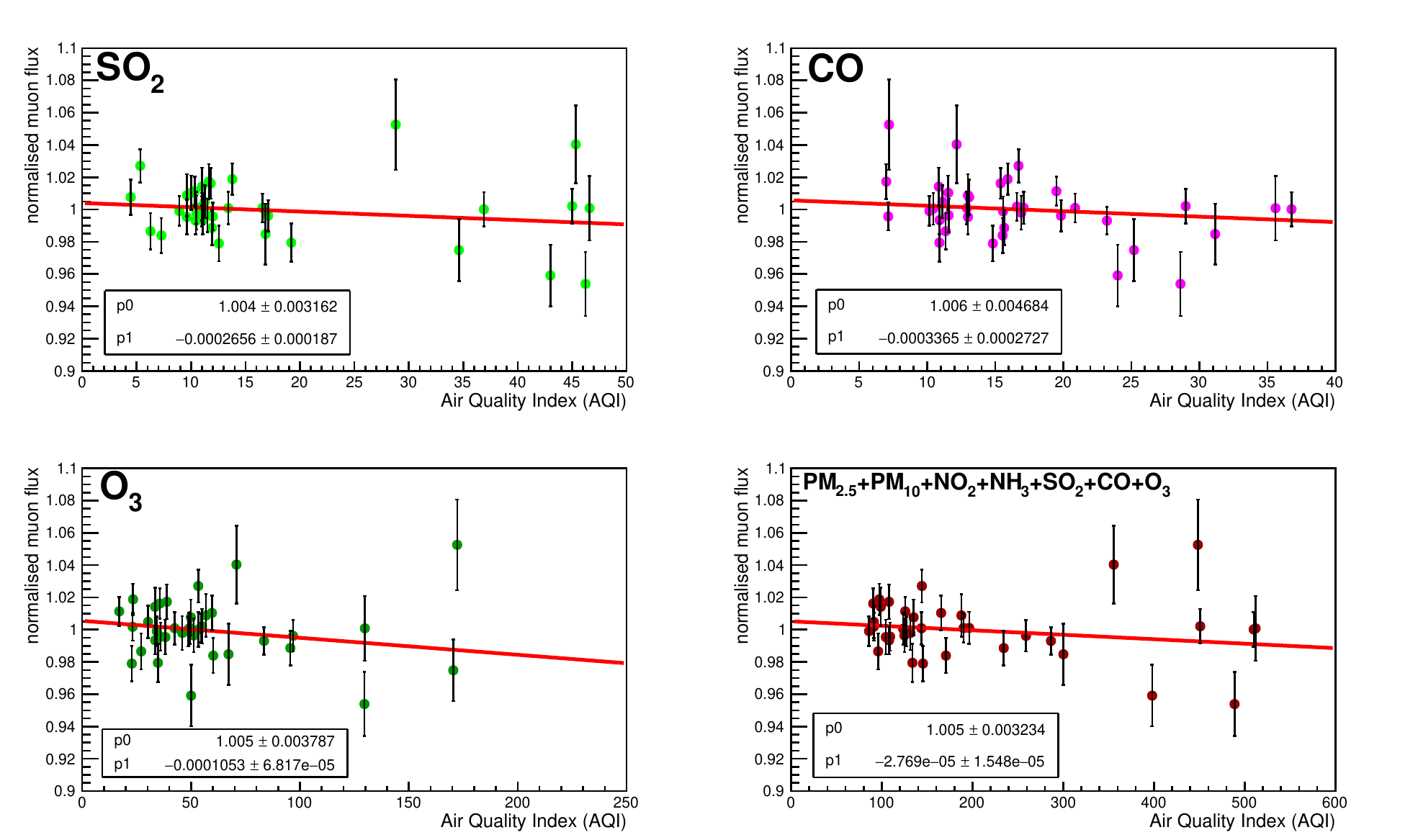}
\captionof{figure}{(Colour online) Normalised muon flux as a function of the air quality index (AQI) of 7 most abundant air pollutants.}
\label{norm_flux_aqi}
\end{figure}

\begin{table} [h!]
\begin{center}
    \begin{tabular}{|c|c|c|}
    \hline
    Air pollutant type & p0 & p1  \\ \hline \hline
    PM$_{2.5}$ & 1.23 $\pm$ 0.003328 & -0.0002962 $\pm$ 5.312e-05 \\ \hline
    PM$_{10}$ & 1.235 $\pm$ 0.003955 & -0.0003795 $\pm$ 6.645e-05  \\ \hline
    NO$_{2}$ & 1.232 $\pm$ 0.003666  & -0.00139 $\pm$ 0.0002634 \\ \hline
    NH$_{3}$ & 1.243 $\pm$ 0.006047  & -0.007628 $\pm$ 0.001701  \\ \hline 
    SO$_{2}$ & 1.236 $\pm$ 0.004197 & -0.001255 $\pm$ 0.0002339 \\ \hline
    CO & 1.243 $\pm$ 0.006191 & -0.001622 $\pm$ 0.0003632 \\ \hline
    O$_{3}$ & 1.226 $\pm$ 0.004799 & -0.0001535 $\pm$ 7.864e-05  \\ \hline 
    Gross & 1.24 $\pm$ 0.004969 & -0.0001563 $\pm$ 3.044e-05  \\
    
\hline
 \end{tabular}
 \caption{Value of the fit parameters of the muon flux vs air quality index (AQI) curve with 7 most abundant air pollutants.}
\label{AP_Table1}
\end{center}
\end{table}

In Fig.~\ref{aqi_date}, we have reported the individual air pollutants (average air quality index (AQI)) before and after the lockdown (on the dates of cosmic ray data recording) and a clear decrement in the concentrations of the air pollutants have been observed after lockdown. One striking thing we observe during this study, is that the concentration of O$_3$ also decreases after lockdown in our case unlike at Ref~\cite{S. Jain} where the concentration of O$_3$ was reported to be increased during lockdown (for a short time period though). Fig.~\ref{flux_aqi} represents the variation of the muon flux with air quality index for the seven air pollutants individually and with the gross pollutants present. We observe a correlation between cosmic ray muon flux and the concentrations of the air pollutants before and after the lockdown where the flux increases with decrease in the concentrations of the air pollutants. The details of the variation as found by linear fitting of the muon flux vs air quality index curve for different pollutants, are tabulated in Table~\ref{AP_Table1} below.

$\it T/p$ normalised muon flux is also plotted as a function of the average air quality index for the seven air pollutants individually and with the gross pollutants present in Fig.~\ref{norm_flux_aqi}. Here also we observed that the normalised flux increases with decrease in the concentrations of the air pollutants.

    

\section{Summary \& Discussion}

Cosmic ray muon flux is measured using the coincidence technique with plastic scintillation detectors. To restrict the outbreak of COVID19, Government of India imposed 67 days of nationwide complete lockdown in three phases. After that, the unlocking was declared in phases in different parts of India. Before the lockdown, we collected some cosmic ray flux data. After lockdown the measurement is continued to compare with the flux as measured before lockdown. In our measurement, it is found that the cosmic ray flux remained more or less unchanged before and after the lockdown. However it is well known that there is an effect of atmospheric temperature and pressure on the cosmic ray flux and we also looked for any such possible correlation. A positive correlation is indeed observed between the muon flux and the ratio of atmospheric temperature and pressure. This correlation is fitted well by a function of the form {\it p0+p1(T/p)}, and the fit parameters {\it p0} and {\it p1} are used to normalise the {\it T/p} effect on the cosmic muon flux. It is found that the mean normalised muon flux before and after the lockdown period are 0.996 with a standard deviation of 0.029 and 1.001 with a standard deviation of 0.012 respectively. 

A lockdown such as the one implemented due to the COVID19 typically has significant influence on the atmospheric condition in terms of presence of pollutants. We wanted to study any possible correlation of measured cosmic ray muon flux with this. To realise this, we considered the seven most abundant air pollutants (PM$_{2.5}$, PM$_{10}$, NO$_{2}$, NH$_{3}$, SO$_{2}$, CO, O$_{3}$) and investigated the change in their concentrations with time (before and after lockdown). We found significant declination in the concentrations of the pollutants and we tried to look for any correlation with the measured muon flux within the stipulated time window. The result shows a clear correlation as with decreasing concentrations of the air pollutants we observed an increasing trend of the normalised muon flux. From our observation, one can comment that the increase in cosmic ray flux can also be considered as one of the secondary indicators of less polluted air.

However, there are a few limitations of our measurement. First, the detector coverage area is very small, resulting in low statistics. Second, the statistics of muon data before lockdown is small. It will be very interesting if any other research laboratory having a large facility of cosmic ray flux measurement can try to study such correlation.
		
\section*{Acknowledgement}
The authors would like to thank Dr. Abhijit  Chatterjee, Prof. Sibaji Raha, Prof.~Rajarshi~Ray, Prof. Somshubhro Bandyopadhyay and Dr. Sidharth K. Prasad for valuable discussions and suggestions in the course of the study. We would also like to thank Mrs. Sharmili Rudra, Dr. Rama Prasad Adak, Mr. Dipanjan Nag, Ms. Nilanjana Nandi and Mr. Subrata Das for helping in the fabrication of the detectors. This work is partially supported by the research grant SR/MF/PS-01/2014-BI from DST, Govt. of India, the research grant of CBM-MuCh project from BI-IFCC, DST, Govt. of India and IRHPA (Intensification of Research in High Priority Areas/Sanction No. IR/S2/PF.01/2011) scheme. A.~Sen acknowledges his Inspire Fellowship research grant [DST/INSPIRE Fellowship/ 2018/IF180361].



\begin{thebibliography}{10}
\bibitem{cosmic_ray}V. Valkovi\`c, Radioactivity in the Environment~(2000), Pages 5-32, https://doi.org/10.1016/B978-044482954-2.50002-2.

\bibitem{S. Jain} S. Jain et al., Aerosol and Air Quality Research, 20: 1222-1236, (2020).
\bibitem{S. Chen} S. Chen et al., https://doi.org/10.1371/journal.pone.0215663 (2019).
\bibitem{A. Chatterjee} Abhijit  Chatterjee et al., Atmospheric Environment,\\ DOI:https://doi.org/10.1016/j.atmosenv.2020.117947.
\bibitem{covid}https://www.covid19india.org/.


\bibitem{pollution}https://app.cpcbccr.com/AQI\_India/.


\bibitem{S. Roy} S. Roy et al., Proceedings of ADNHEAP 2017, Springer Proceedings in Physics 201, 199 - 204, ISBN 978-981-10-7664-0.
\bibitem{S. Shaw} S. Shaw et al., Proceedings of the DAE Symp. on Nucl. Phys. 62 (2017) 1038.
\bibitem{M. Neira} M. Neira, A. Pr\"uss-Ust\"un and P. Mudu, Lancet 392: 1178-1179 (2018).
\bibitem{M. Zazyan} M. Zazyan et al., J. Space Weather Space Clim., 5, A6 (2015).
\bibitem{tp}https://www.timeanddate.com/weather/india/kolkata

			

			
\end{thebibliography}
\end{document}